# Electro-Fenton treatment of benzophenone-4 solutions: A sustainable approach for its removal using an air-diffusion cathode

Caio Machado Fernandes[a,b,**], Enric Brillas[c], Mauro C. Santos[a], Sergi Garcia-Segura[b,*]

[a] *Laboratório de Eletroquímica e Materiais Nanoestruturados, Centro de Ciências Naturais e Humanas, Universidade Federal do ABC, Santo André, SP, 09210-170, Brasil*

[b] *Nanosystems Engineering Research Center for Nanotechnology-Enabled Water Treatment, School of Sustainable Engineering and the Built Environment, Arizona State University, Tempe, AZ, 85287, USA*

[c] *Departament de Ciència de Materials i Química Física, Facultat de Química, Universitat de Barcelona, Martí i Franqus 1-11, 08028, Barcelona, Spain*

Corresponding Author: *E-mail: sgarcias@asu.edu (S. Garcia-Segura)

\*\*E-mail: cmacha13@asu.edu (C. Machado Fernandes)

**Abstract**

This work reports the efficient degradation and mineralization of benzophenone-4 (BP-4), a widely used UV filter associated with endocrine-disrupting effects, via the electro-Fenton process. Key operating parameters including pH, current density, $Fe^{2+}$ dosage, and initial pollutant concentration were optimized. At pH = 3.0, the best performance was obtained. Under optimum conditions of 20 mA cm$^{-2}$ and 0.75 mM $Fe^{2+}$, complete BP-4 degradation was achieved in remarkably short times: 2 min for 1–2.5 mg L$^{-1}$, 3 min for 5 mg L$^{-1}$, 4 min for 10 mg L$^{-1}$, 5 min for 20 mg L$^{-1}$, and 7 min for 40 mg L$^{-1}$. Total mineralization was reached for low concentration of 1 mg L$^{-1}$, varying between 77 and 90% for higher loads up to 40 mg L$^{-1}$. The reaction followed pseudo-first-order kinetics, with high apparent rate constant of 0.641 min$^{-1}$ for 40 mg L$^{-1}$ and lower energy consumption of 0.261 kWh (g TOC)$^{-1}$. Radical quenching experiments with TBA confirmed that physisorbed BDD($^{\bullet}$OH) and homogeneous •OH were the predominant oxidants. Persistent carboxylic acid by-products, in the form of $Fe^{3+}$ complexes, were the only residues after 180 min of treatment, readily being biodegradable. The process demonstrates high efficiency across a broad range of BP-4 concentrations, offering a viable solution for its removal from contaminated waters.

# 1. Introduction

Emerging contaminants including pharmaceuticals and personal care products (PPCPs) have raised significant environmental concerns due to their continuous release into aquatic ecosystems and their persistence in the environment (Felisardo et al., 2024). Among them, ultraviolet (UV) filters, commonly found in sunscreens and cosmetics, have garnered increased attention in recent years due to their widespread daily use and their high potential to reach aquatic systems through multiple sources (Esperanza et al., 2019; Vuckovic et al., 2023; Wu et al., 2018).

Benzophenone-4 (BP-4), also known as sulisobenzone (5-Benzoyl-4-hydroxy-2-methoxybenzenesulfonic acid, $C_{14}H_{12}O_6S$, $M$ = 308.31 g mol$^{-1}$, see chemical structure in Fig. 1) ranks among the most typically employed UV filters within the benzophenone class due to its effective skin protection against harmful radiation (Ruz-Luna et al., 2024). The BP-4 is extensively applied to sunscreens and cosmetics, but it is also found in plastics and other products to enhance polymer resistance to weathering (Santbay et al., 2024). Its widespread application has led to frequent detection in environmental matrices and reported by numerous research conducted worldwide (Amankwah et al., 2024; Bratkovics et al., 2015; Kasprzyk-Hordern et al., 2008; O'Malley et al., 2020). Monitoring efforts have revealed the presence of BP-4 in various aquatic environments, indicating its persistence and the limitations of conventional wastewater treatment processes in removing it. Several studies have also shown that BP-4 poses a serious risk to both aquatic ecosystems and human health, being linked to endocrine disruption (Kunz and Fent, 2006; Liu et al., 2016; Zucchi et al., 2011), toxicity (Du et al., 2017; Liu et al., 2015; Paredes et al., 2014; Tao et al., 2024; Yang et al., 2024; Zhang et al., 2017) and potential bioaccumulation in organisms (Carve et al., 2021; Ruszkiewicz et al., 2017). Moreover, concerns are growing over human exposure through contaminated water sources (Hopkins and Blaney, 2016), reporting BP-4 accumulated in the placenta (Valle-Sistac et al., 2016) and even detection in the urine of some individuals (Kim et al., 2022). Given these environmental and health risks, the need for effective removal strategies of this pollutant is becoming more urgent.

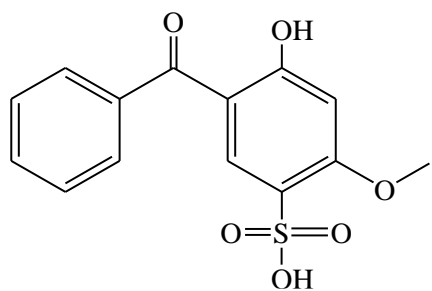

**Fig. 1.** Chemical structure of BP-4.

Wastewater treatment plants (WWTPs) have demonstrated limited efficacy in destroying emerging contaminants like BP-4. Personal care products such as BP-4, along with other UV filters, tend to resist conventional WWTP treatments and may even pass through the entire treatment train largely unaltered, ultimately accumulating in aquatic environments. This inefficacy stems from the chemical stability and low biodegradability of BP-4, allowing it to persist in treated effluents, surface waters, and sediments, thus contributing to its continuous environmental cycling (De Laurentiis et al., 2013; Semones et al., 2017). In light of these limitations, the electrochemical advanced oxidation processes (EAOPs) can help to solve this pollution problem if they were to be implemented as tertiary treatments in WWTPs targeting high-performance degradation of resistant pollutants (Brillas, 2025; Can-Güven et al., 2025; Ganiyu et al., 2021; Garcia-Segura and Brillas, 2024; Magdaleno et al, 2024).

The EAOPs are highly effective methods that rely on the *in-situ* generation of reactive oxygen species (ROS) including the hydroxyl radical (•OH, $E^o$ = 2,80 V (vs. SHE)). •OH is the second strongest oxidant known after fluorine, capable of completely degrading and mineralizing most organic aromatic pollutants (Zhu et al., 2025; Garcia-Segura et al., 2018). Among EAOPs, electrochemical oxidation (ECO) is one of the most extensively applied. In ECO, the organic compounds are broken down by physiosorbed hydroxyl radicals (M(•OH)) produced at an anode material (M) with a high $O_2$ evolution overpotential during water oxidation according to Eq. (1) (Wang et al., 2023).

Electrochemically driven Fenton processes represent a promising alternative with higher oxidation power than ECO treatment. These emerging electrified Fenton technologies combine the action of heterogeneous M(•OH)) generated at the anode surface with homogeneous •OH produced in the bulk solution through Fenton reaction between electrogenerated $H_2O_2$ and added $Fe^{2+}$, expressed as Eq. (2). In the electro-Fenton (EF) process, hydrogen peroxide acts as a key reactant, being cathodically generated on site via oxygen reduction reaction by Eq. (3)). Additionally, the electrochemical system promotes the continuous regeneration of the $Fe^{2+}$ catalyst from $Fe^{3+}$ reduction by Eq. (4). This rapid electro-regeneration of the catalyst drastically decreases the required $Fe^{2+}$ content compared to the traditional chemical Fenton process, enhancing both the sustainability and cost-effectiveness of the treatment (Brillas et al., 2009; Huo et al., 2025; Liang et al., 2025; Nienhauser et al., 2022; Yang et al., 2023; Yu et al., 2025). In EF, the main oxidants are M(•OH) formed at the anode from reaction (1) and •OH from Fenton reaction (2). In several studies and for comparative purposes, the ECO process is additionally performed with $H_2O_2$ electrogeneration, without $Fe^{2+}$ addition (also known as ECO-$H_2O_2$).

$$M + H_2O \longrightarrow M(\bullet OH) + H^+ + e^- \tag{1}$$

$$H_2O_2 + Fe^{2+} \rightarrow Fe^{3+} + \bullet OH + {}^-OH \tag{2}$$

$$O_2 + 2H^+ + 2e^- \rightarrow H_2O_2 \tag{3}$$

$$Fe^{3+} + e^- \rightarrow Fe^{2+} \tag{4}$$

To gain a better understanding of EF processes for water remediation, this work presents the first systematic study usig a BDD/air-diffusion system for the complete mineralization of BP-4, a persistent and ecotoxic UV filter. The comprehensive assessment of this electrochemical approach is then shown as a robust strategy for the removal of this pollutant. The goal of this work is to assess and optimize the efficiency of BP-4 removal under several operating variables like pH, current density (*j*), and $Fe^{2+}$ and pollutant concentrations in order to provide an enhanced approach for its environmental remediation. The final carboxylic acids were identified and their impact on the mineralization process was analyzed, and key figures of merit were evaluated.

## 2. Materials and methods

*2.1. Chemicals*

Analytical grade BP-4, $Na_2SO_4$, $FeSO_4 \cdot 7H_2O$, and carboxylic acids were bought from Sigma-Millipore. Ultrapure water from an Elga Water system was employed for preparing all solutions. Analytical grade $H_2SO_4$ and NaOH used for pH adjustment were purchased from Sigma-Millipore. All additional chemicals were of either analytical or high-performance liquid chromatography (HPLC) grade.

*2.2. Electrochemical systems*

All electrochemical measurements were conducted with a standard 250 mL undivided cell containing 150 mL of the test solution at room temperature (about 25 ºC). To maintain homogeneity with mass transport enough of the contaminant towards/from the electrode surfaces, the solution was continuously stirred at 300 rpm. A BDD electrode of 10 cm² provided from NeoCoat was used as the anode, whereas the cathode was a PTFE-treated carbon cloth of 2.5 cm² acquired from the Fuel Cell Store (College Station, TX, USA) was assembled as an air-diffusion electrode, which enabled continuous $H_2O_2$ production via reaction (2). The solution was bubbled with air with a pump at 2.0 L min$^{-1}$ flow rate.

All solutions were prepared with 0.050 M $Na_2SO_4$ as supporting electrolyte. The influence of crucial operating parameters was assessed by varying the pH (1.0–7.0), $Fe^{2+}$ catalyst dosage (0–1.00 mM), and BP-4 concentration (1.0–40 mg L$^{-1}$). The pH was adjusted using 1 M $H_2SO_4$. Galvanostatic assays were made at a constant $j$ relative to the cathode, varying from 5 to 40 mA cm$^{-2}$ (currents of 12.5 to 100 mA) provided by a TENMA 72–2720 DC potentiostat/galvanostat as a power source. During the experiments, 1. 5 and 30 mL aliquots were collected at predetermined intervals for subsequent analysis of BP-4 degradation and total organic carbon (TOC), respectively. To ensure reproducibility and reliability, all experiments were conducted in duplicate, and figures include the error bars representing a 95% confidence interval. Before each electrolysis run, the cleaning and activation of the electrodes were achieved by electrolyzing them in a 0.050 M $Na_2SO_4$ solution at $j$ = 100 mA cm$^{-2}$.

*2.3. Analytical methods and calculations*

Quantification of electrogenerated $H_2O_2$ was performed using a colorimetric method, where the yellow Ti(IV)–$H_2O_2$ complex was formed and its absorbance measured at $\lambda$ = 408 nm with a DR6000 UV–Vis spectrophotometer (Hach, Ames, IA, USA). Faradaic efficiency (%) for $H_2O_2$ accumulation was calculated based on Faraday's law from Eq. (5):

$$\% \text{ Faradaic efficiency} = \frac{[H_2O_2] \, n \, F \, V_s}{I \, t \, M} \, 100 \tag{5}$$

where [$H_2O_2$] is the measured $H_2O_2$ concentration (g L$^{-1}$), $n$ = 2 (number of electrons transferred from reaction (3)), $F$ is Faraday constant (96,487 C mol$^{-1}$), $V_s$ is the solution volume (L), $I$ is the applied current (A), $t$ is the electrolysis time (s), and $M$ is the molar mass of $H_2O_2$ (34 g mol$^{-1}$).

The BP-4 decay over electrolyzed time was monitored with a Waters 2695 HPLC equipped with a Waters Symmetry C18 column (4.6 mm × 75 mm, 3.5 µm) and connected to a Waters 2998 photodiode array detector set to $\lambda$ = 278 nm. A sample of 20 µL was injected into the LC and eluted with 20:80 (v/v) acidified water/methanol mixture at 0.5 mL min$^{-1}$ flow rate. The BP-4 peak appeared in the chromatogram at a retention time of 3.5 min, and no additional peaks corresponding to by-products were observed.

The degradation kinetics of BP-4 was analyzed with a pseudo-first-order kinetic reaction expressed from Eq. (6):

$$\ln\left(\frac{c_0}{c}\right) = k_1 \, t \tag{6}$$

where $c_0$ and $c$ represent the initial concentration of BP-4 and its concentration at time $t$, and $k_1$ is the pseudo-first-order rate constant. For each trial, Eq. (6) was applied using the KaleidaGraph 4.03 program and it was acceptable for a square regression coefficient ($R^2$) ≥ 0.980.

The electrical energy per order (EE/O, in kWh m$^{-3}$ order$^{-1}$), meaning the energy needed to reduce the BP-4 content by one order of magnitude in a unit volume, was calculated from Eq. (7) (dos Santos et al., 2021):

$$\text{EE/O} = \frac{E_{\text{cell}} \, I \, t}{V_s \log\left(\frac{c_0}{c}\right)} = \frac{E_{\text{cell}} \, I}{0.4343 \, V_s k_1} \tag{7}$$

where the average cell potential (in V) is symbolized as $E_{\text{cell}}$, the current (in A) by $I$, the time (in h) by $t$, the solution volume (in L) by $V_s$, and $k_1$ is given in h$^{-1}$. The second term of EE/O can be applied when a pseudo-first order kinetics (log ($c_0/c$) = 0.4343 $k_1$ $t$) is verified.

The same HPLC was used to quantify the carboxylic acids, but equipped with a Bio-Rad Aminex HPX-87H column (7.8 mm × 300 mm) maintained at 35 °C and the detector chosen at λ = 210 nm. A volume of 20 μL of sample was also injected into the LC that eluted with 4 mM $H_2SO_4$ at 1 mL min$^{-1}$ flow rate. The retention times were of 6.1 min for oxalic acid, 7.5 min for acetic acid, 13.8 min for formic acid, and 18.0 min for fumaric acid. All the assays were made by duplicate and the average values of concentrations determined are reported, with the error bars related to a 95% confidence interval shown in the figures-of-merit.

The organic content in the solution was quantified through TOC measurements with a Shimadzu TOC-L carbon analyzer by injecting 30 mL samples. Triplicate TOC measurements were made for each sample with the non-purgable organic carbon method with an error < 1%. The mineralization current efficiency (MCE, in %) and the energy consumption (EC$_{\text{TOC}}$, in kWh (g TOC)$^{-1}$) were calculated from Eq. (8) and (9), respectively (Brillas et al., 2009; Hu et al., 2024):

$$\text{MCE} = \frac{n \, F \, V_s \, (\text{TOC}_0 - \text{TOC})}{4.32 \times 10^7 \, m \, I \, t} \times 100 \tag{8}$$

$$\text{EC}_{\text{TOC}} = \frac{E_{\text{cell}} \, I \, t}{V_{\text{s}}(\text{TOC}_{\text{o}} - \text{TOC})} \tag{9}$$

where $n$ represents the number of electrons in the complete mineralization of a BP-4 molecule, $F$ is the Faraday constant (96,485 C mol$^{-1}$), TOC$_0$ and TOC are the initial total organic carbon and that at time $t$, $m$ is the number of carbon atoms of BP-4, and $t$ is the time (in s for Eq. (8) and in h for Eq. (9)). The $n$-value was assumed to be 62 according to Eq. (10) assuming that the sulfur atom (S) is transformed into SO$_4^{2-}$ ion:

$$C_{14}H_{12}O_6S + 26H_2O \rightarrow 14CO_2 + SO_4^{2-} + 64H^+ + 62e^- \tag{10}$$

## 3. Results and discussion

*3.1. Understanding the electrogeneration of hydrogen peroxide.*

Fig. 2a presents the accumulation profile of H$_2$O$_2$ over time in 0.050 M Na$_2$SO$_4$ at pH =3.0 using the BDD/ air-diffusion system at $j$ values of 5-40 mA cm$^{-2}$. After 60 min of electrolysis, H$_2$O$_2$ concentrations reached 55.2, 92.6, 181.7, and 300.5 mg L$^{-1}$ for 5, 10, 20, and 40 mA cm$^{-2}$, respectively, increasing further to 67.3, 112.6, 221.4, and 381.7 mg L$^{-1}$ at 90 min. This rise in H$_2$O$_2$ yield with both, time and $j$ reflected that the enhanced electron transfer rate by Eq. (3) maintained the optimum O$_2$ availability at the reaction interface. Fig. 2b shows tthe Faradaic efficiency for H$_2$O$_2$ accumulation calculated from Eq.(5) for the above assays. They achieved high values near 100% at the beginning of the electrolysis at $j$ = 5 mA cm$^{-2}$ that slowly decayed to 85% at 90 min. The same tendency was found for the other trials with starting efficiencies decreasing up to 76% for $j$ =40 mA cm$^{-2}$. This behavior indicates that reaction (3) is the main cathodic process and the H$_2$O$_2$ decay is due to its oxidation at the anode via reaction (11) originating the weak oxidant hydroperoxyl radical (HO$_2^\bullet$) (Brillas et al., 2009). Such sustained *in situ* H$_2$O$_2$ production is critical for efficient $^\bullet$OH generation by Fenton reaction (2), directly linking oxidant accumulation to the degradation capacity of the system in subsequent homogeneous EF treatment ((Cordeiro-Junior et al., 2022: Zhao et al., 2023).

$$\text{M} + \text{H}_2\text{O}_2 \rightarrow \text{M}(\text{HO}_2^\bullet) + \text{H}^+ + e^- \tag{11}$$

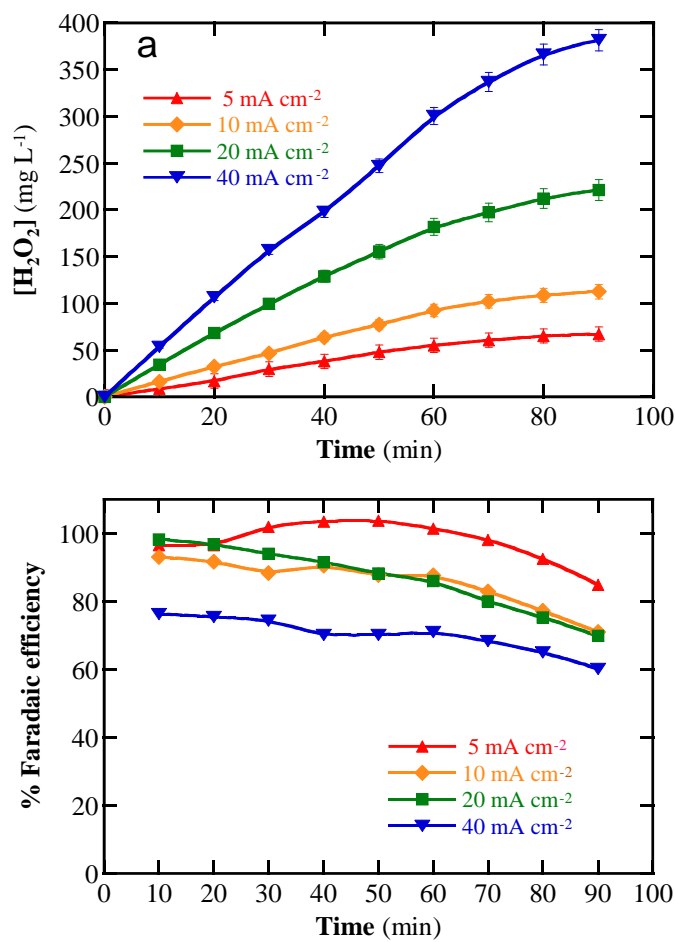

**Fig. 2.** Effect of current density on the time course of (a) the concentration of accumulated $H_2O_2$ and (b) Faradaic efficiency for 150 mL of a 0.050 M $Na_2SO_4$ at pH = 3.0 using a stirred undivided tank reactor with a BDD anode and a carbon cloth air-diffusion cathode.

*3.2. Elucidating the critical role of pH*

The first operating variable assessed was the solution pH because it largely influences the rate of Fenton reaction (2) to generate $^\bullet OH$ and therefore impacts the oxidation power of the EF treatment (Brillas et al., 2009; Ganiyu et al. 2021). This effect was studied in experiments with 10 mg L$^{-1}$ of BP-4 and 0.50 mM Fe$^{2+}$ catalyst at $j$ = 20 mA cm$^{-2}$. Fig. 3a shows that the normalized pollutant concentration under such conditions decreased when the pH increased from 1.0 to 3.0, where its value became zero after 5 min of electrolysis. The lower oxidation ability at pH = 1.0 can be ascribed to the partial protonation of electrogenerated $H_2O_2$ from Eq. (3) to originate the $H_3O_2^+$ cation, which was not susceptible to reacting with the $Fe^{2+}$ catalyst (Ganiyu et al. 2021). In contrast, at pH = 3.0, all electrogenerated $H_2O_2$ is expected to be quickly consumed by such a catalyst and hence, the Fenton reaction (2) achieved its higher rate with maximum •OH production. This hypothesis was confirmed by the loss of reactivity of the system at pH > 3.0. So, at 5 min of treatment, the BP-4 degradation was diminished from 100% to 88% at pH = 5.0, and much more drastically down to solely 23% at pH = 7.0. Such diminished effectiveness stems primarily from the gradual decrease of soluble $Fe^{2+}$ with increasing pH because of the precipitation of iron hydroxides including $Fe(OH)_2$ and $Fe(OH)_3$ from the $Fe^{3+}$ formed from reaction (2) (Brillas et al., 2009; Zhuo et al., 2023). The optimum pH for BP-4 degradation was then reached at pH 3.0, close to that of reaction (2) (Brillas, 2023; Mbaye et al., 2022), meaning that sustaining efficient •OH production is essential. This pH was then chosen to assess the effect of the other operating variables in the next experiments.

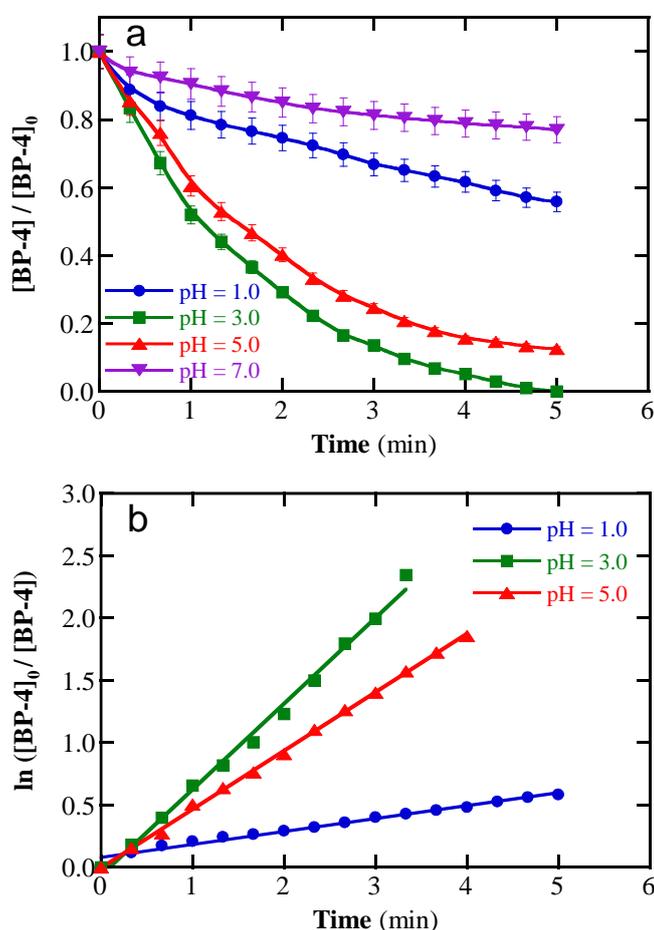

**Fig. 3**. Effect of pH on (a) the normalized BP-4 concentration vs. time and (b) the corresponding pseudo-first-order kinetic analysis for the homogeneous EF process of 150 mL of 10 mg $L^{-1}$ pollutant in ultra-pure water with 0.050 M $Na_2SO_4$ and 0.50 mM $Fe^{2+}$ using a stirred undivided BDD/air-diffusion tank reactor at cathodic current density ($j$) of 20 mA $cm^{-2}$.

The concentration of the above experiments was further analyzed with a pseudo-first-order reaction for BP-4 degradation. Fig. 3b depicts excellent linear correlations obtained up to pH = 5.0, except at pH = 7.0 probably by an irregular •OH production associated to the progressive precipitation of $Fe(OH)_3$. The calculated $k_1$-values were 0.104 min$^{-1}$ ($R^2$ = 0.980) for pH = 1.0, 0.689 min$^{-1}$ ($R^2$ = 0.994) for pH = 3.0, and 0.469 min$^{-1}$ ($R^2$ = 0.998) for pH = 5.0. This presupposes that a steady concentration of the oxidizing species BDD(•OH) and •OH is produced at each pH value from reactions (1) and (2), respectively. The relevance of pH as a master variable that defines $H_2O_2$ and $Fe^{2+}$ speciation in solution becomes a main driver of degradation performance, which achieves its maximum value at the optimum pH = 3.0.

*3.3. Balancing removal kinetics with process efficiency through the control of cathodic current density*

The cathodic *j* determines the rate of charge transfer reactions, thus being a key factor for the oxidation ability of the EF process. The effect of this operating variable was assessed from the removal of solutions with 10 mg L$^{-1}$ BP-4 and 0.50 mM $Fe^{2+}$ at pH = 3.0, and the experimental findings are displayed in Fig. 4a. As can be seen, overall degradation was reached at 20 min for *j* = 5 mA cm$^{-2}$, decreasing at 10 min for *j* = 10 mA cm$^{-2}$, 5 min for *j* = 20 mA cm$^{-2}$, and 3 min for *j* = 40 mA cm$^{-2}$. Fig. 4b highlights the outstanding linear fitting from a pseudo-first-order kinetics with $k_1$-values listed in Table 1. The $k_1$ progressively raised from 0.231 to 1.384 min$^{-1}$ when changing from 5 to 40 mA cm$^{-2}$, higher than other reported studies (Machado Fernandes et al., 2025, Liang et al., 2025, Dang et al., 2025). The observed trend is attributable to enhanced rate of reactions (1) and (3) that results in the production of more amounts of BDD(•OH) and $H_2O_2$, respectively. Greater $H_2O_2$ production then enhanced •OH from Fenton reaction (2). Nevertheless, these $k_1$-values indicate that the rate of the pollutant degradation increased 6.0 times when $j$ was 8-fold higher. This loss of reactivity can be associated with a concomitant acceleration of undesired non-oxidizing parasitic reactions of the produced ROS as $j$ increased. Examples of such undesired reactions are reactions (12) and (13) involving $^{\bullet}OH$ dimerization and its attack with $H_2O_2$ yielding the weaker oxidant hydroperoxyl radical ($HO_2^{\bullet}$) (Brillas et al., 2009).

$$2^{\bullet}OH \rightarrow H_2O_2 \tag{12}$$

$$^{\bullet}OH + H_2O_2 \rightarrow HO_2^{\bullet} + H_2O \tag{13}$$

From the aforementioned findings, other indicators different from $k_1$ were analyzed to ascertain the best $j$-value to be applied. The EE/O values determined for the above experiments also gradually grew from 0.0343 to 0.1614 kWh m$^{-3}$ order$^{-1}$, as listed in Table 1. This increase was also due to the mg L$^{-1}$ pollutant in ultra-pure water with 0.050 M Na$_2$SO$_4$ and 0.50 mM Fe$^{2+}$ at pH = 3.0 using a stirred undivided BDD/air-diffusion tank reactor.

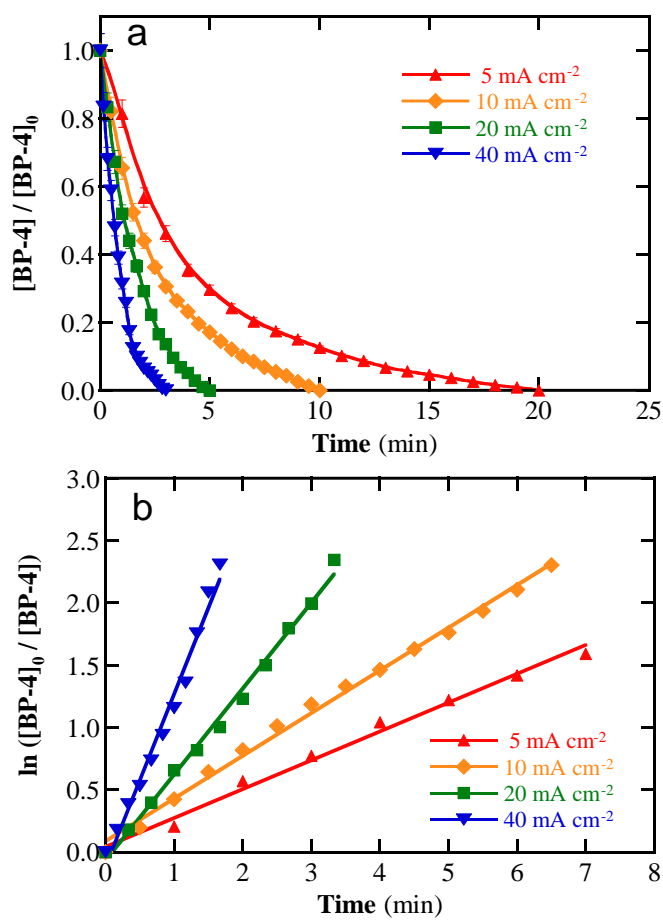

**Fig. 4.** Influence of cathodic current density on (a) the change of normalized BP-4 concentration with time and (b) the corresponding pseudo-first-order kinetic analysis for the homogeneous EF treatment of 150 mL of 10

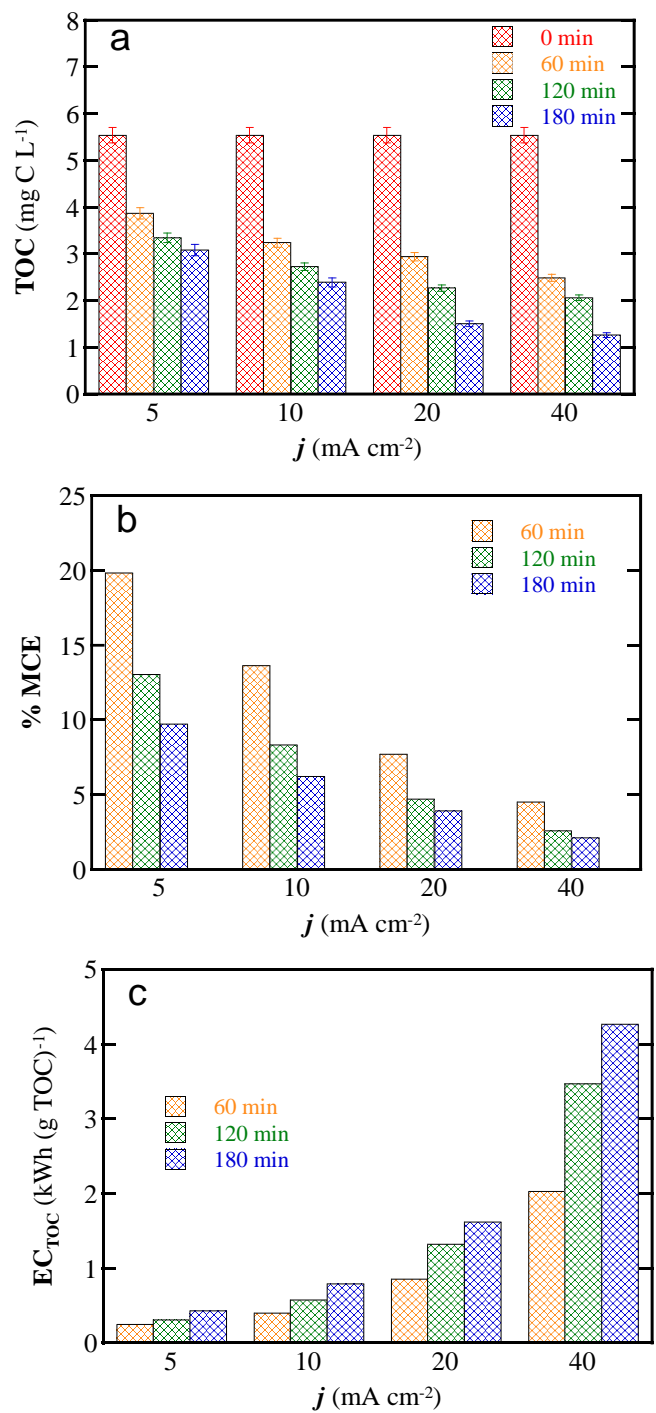

**Fig. 5.** Effect of cathodic current density on the time course of (a) TOC, (b) mineralization current efficiency, and (c) energy consumption per unit TOC mass for the homogeneous EF treatment of 150 mL of 10 mg L$^{-1}$ pollutant in ultra-pure water containing 0.050 M Na$_2$SO$_4$ and 0.50 mM Fe$^{2+}$ at pH = 3.0 using a stirred undivided BDD/air-diffusion tank reactor.

**Table 1.** Pseudo-first-order rate constant ($k_1$) for BP-4 degradation with the square regression coefficient, electrical energy per order (EE/O), and percent of total organic carbon (TOC) and mineralization current efficiency (MCE) with energy consumption per unit TOC mass ($EC_{TOC}$) after 180 min of electrolysis of 150 mL of pollutant solutions in ultra-pure water with 0.050 M $Na_2SO_4$ under different experimental conditions using a stirred undivided BDD/air-diffusion cell.

| [BP-4] (mg L$^{-1}$) | $j$ (mA cm$^{-2}$) | [Fe$^{2+}$] (mM) | $k_1$ (min$^{-1}$) | $R^2$ | EE/O (kWh m$^{-3}$ order$^{-1}$) | % TOC removal | % MCE | $EC_{TOC}$ (kWh (g TOC)$^{-1}$) |
|---|---|---|---|---|---|---|---|---|
| 1.0 | 20 | 0.75 | 2.268 | 0.992 | 0.0288 | 100 | 0.5 | 10.847 |
| 2.5 | 20 | 0.75 | 2.141 | 0.980 | 0.0305 | 90 | 0.9 | 5.715 |
| 5.0 | 20 | 0.75 | 1.503 | 0.990 | 0.0435 | 79 | 1.9 | 2.919 |
| 10 | 5 | 0.50 | 0.231 | 0.990 | 0.0343 | 44 | 9.7 | 0.432 |
| 10 | 10 | 0.50 | 0.342 | 0.995 | 0.0719 | 57 | 6.2 | 0.791 |
| 10 | 20 | 0 | 0.095 | 0.991 | 1.2180 | 43 | 2.4 | 3.931 |
| 10 | 20 | 0.25 | 0.502 | 0.986 | 0.1797 | 55 | 3.0 | 2.472 |
| 10 | 20 | 0.50 | 0.689 | 0.994 | 0.1057 | 73 | 3.9 | 1.619 |
| 10 | 20 | 0.75 | 1.261 | 0.998 | 0.0522 | 77 | 4.2 | 1.374 |
| 10 | 20 | 1.00 | 1.021 | 0.994 | 0.0554 | 75 | 4.1 | 1.240 |
| 10 | 40 | 0.50 | 1.384 | 0.983 | 0.1614 | 77 | 2.1 | 4.264 |
| 20 | 20 | 0.75 | 0.905 | 0.996 | 0.0723 | 79 | 8.9 | 0.750 |
| 40 | 20 | 0.75 | 0.641 | 0.991 | 0.1021 | 88 | 19.4 | 0.441 |

increase of $E_{cell}$ with $j$, with values of 2.48, 3.84, 5.70, and 8.74 V for 5, 10, 20, and 40 mA cm$^{-2}$, respectively. The EE/O increase allows inferring greater operational expenditures at higher applied $j$, although with much rapid degradation. To achieve more significant information about the mineralization potential of the EF process, the TOC decay under the above conditions was monitored.

Fig. 5a illustrates the TOC abatement and the overall percentage of TOC removed after 180 min is shown in Table 1. A gradual drop of TOC with time can always be observed, as expected from the mineralization of the intermediates formed. The decrease in TOC significantly raised when increasing $j$ from 5 to 20 mA cm$^{-2}$ the percentage of TOC abatement grew from a discrete 44% at $j$ = 5 mA cm$^{-2}$ to 73% at $j$ = 20 mA cm$^{-2}$. However, further increases in $j$ only raised up to 77% of TOC removal at $j$ = 40 mA cm$^{-2}$. This strong inhibition after going from 20 to 40 mA cm$^{-2}$ suggests that most of the extra

ROS generated at the higher $j$ are rather consumed by parasitic reactions due to the recalcitrance of by-products formed, suggesting that $j = 20$ mA cm$^{-2}$ is the best option to operate EF treatment. The MCE and EC$_{TOC}$ values are depicted in Fig. 5b and 5c, respectively. In all cases, the MCE value decayed with time for two main reasons: the reduction in organic content coupled with the generation of persistent by-products that hinder overall mineralization (Huang et al., 2021). Moreover, this parameter also decreased with raising $j$, as expected by the enhanced kinetics of undesirable side processes, shown in Table 1 for 180 min of electrolysis. Conversely, the EC$_{TOC}$ values increased with increasing time and $j$, in agreement with the corresponding loss of reactivity of the EF process. Note that in such values, the electric contribution of the air pump of 10$^{-3}$ $t$ (min) in kWh (g TOC$^{-1}$) was added to that of Eq. (9). This contribution was insignificant as compared to the energy provided to the cell. From all these considerations, $j = 20$ mA cm$^{-2}$ served as the basis for follow-up studies because it provided a very fast pollutant degradation with a relatively high TOC abatement at 180 min. Also, it is interesting to highlight that those energy consumption and electrical energy per order values were lower than reported for pollutant degradation with different techniques beside EF process (Brienza et al., 2022; Castro-Fernandez et al., 2025; Chen et al., 2025; Dominguez et al., 2018; Escalona-Durán et al., 2025; Garcia et al., 2019; Zhou et al., 2019).

*3.4. Identifying the optimum dosage of Fe$^{2+}$ as Fenton catalyst*

Another key operating variable in EF is the dosage of Fe$^{2+}$ catalyst to act over the Fenton reaction (2) and consequently control •OH formation. Fig. 6a reveals the effect of this parameter on the normalized BP-4 content vs. time for 10 mg L$^{-1}$ of this pollutant at pH = 3.0 and $j = 20$ mA cm$^{-2}$. When no Fe$^{2+}$ was present in the solution, corresponding to the ECO-H$_2$O$_2$, a poor degradation of 39% was attained at 5 min mainly due to the slow reactivity of produced BDD(•OH) via reaction (1) as main oxidant in the system. Note that because the heterogeneous character of physiosorbed BDD(•OH), the mass transfer limits the BP-4 degradation by from/towards the BDD electrode. The addition of Fe$^{2+}$ strongly accelerated the BP-4 removal by the oxidation with •OH produced from Fenton reaction (2), more efficient than the parallel oxidation with BDD(•OH). For 0.25 mM $Fe^{2+}$, 93% degradation was reached, whereas total removal was achieved for 0.50 mM $Fe^{2+}$ as a result of a higher rise in the rate of the Fenton reaction (2) induced by a higher catalyst dose. The same behavior can be observed when $Fe^{2+}$ is raised to 0.75 mM, where the pollutant was completely abated in a shorter treatment time of 4 min, meaning an enhancement of •OH generation. In contrast, a further increase to 1.00 mM $Fe^{2+}$ hindered the BP-4 decay, and total removal was obtained after 4.7 min of electrolysis. This can be associated with the greater consumption of the extra generated •OH with the higher amount of $Fe^{2+}$ via the parasitic reaction (14) (Zhang et al., 2016). From these findings, one can conclude that 0.75 mM $Fe^{2+}$ is optimal under the present experimental conditions and this dose was selected for further tests.

$$Fe^{2+} + {}^{\bullet}OH \rightarrow Fe^{3+} + OH^- \tag{14}$$

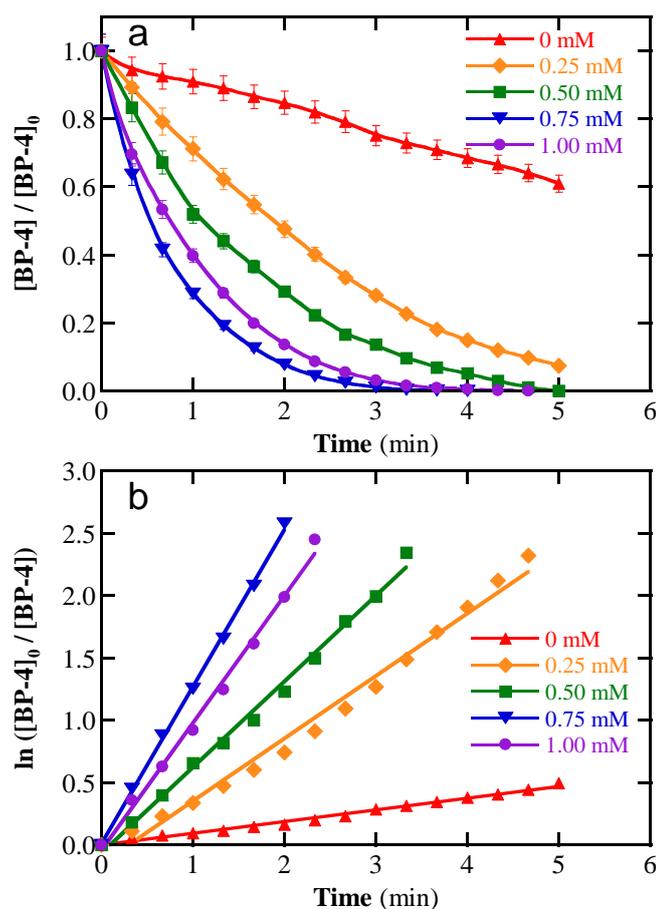

**Fig. 6.** Effect of $Fe^{2+}$ dosage on (a) the variation of normalized BP-4 concentration with time and (b) its pseudo-first-order kinetic analysis for the EF treatment of 150 mL of 10 mg $L^{-1}$ pollutant in ultra-pure water with 0.050 M $Na_2SO_4$ at pH = 3.0 using a stirred undivided BDD/air-diffusion tank reactor at $j$ = 20 mA $cm^{-2}$.

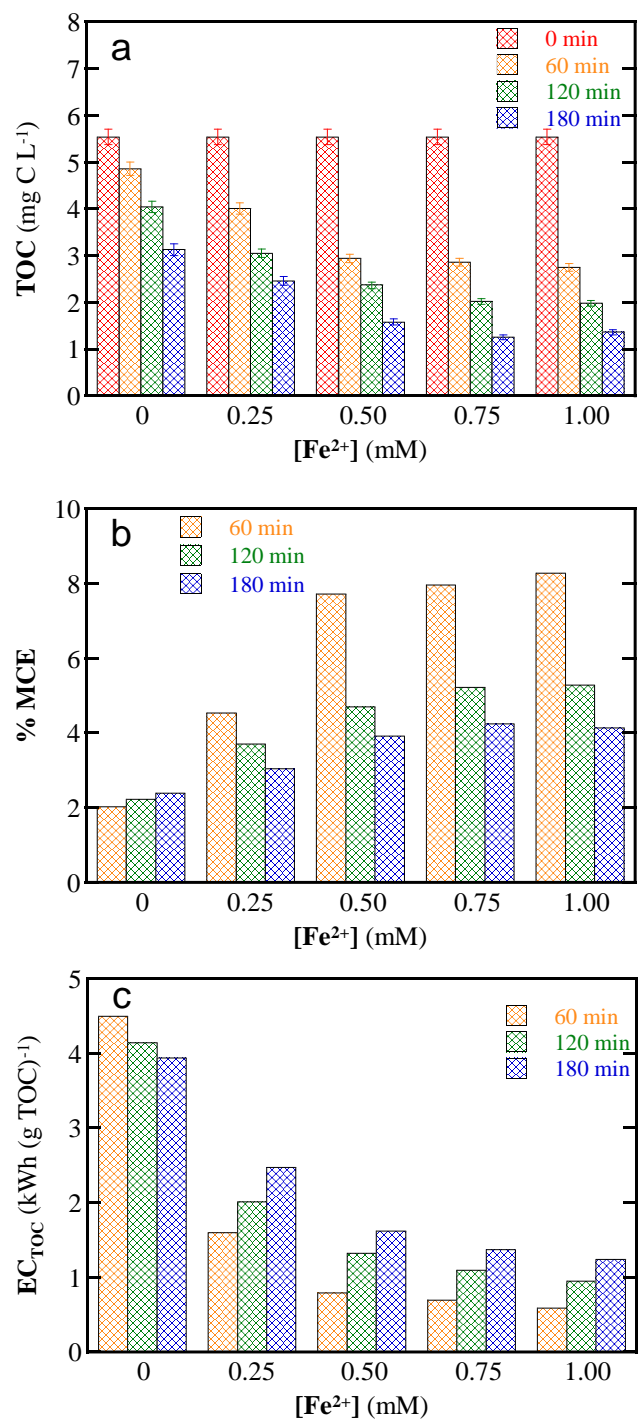

**Fig. 7.** Influence of $Fe^{2+}$ dosage on the change of (a) TOC, (b) MCE, and (c) $EC_{TOC}$ with time for the EF treatment of 150 mL of 10 mg L$^{-1}$ pollutant in ultra-pure water with 0.050 M $Na_2SO_4$ at pH = 3.0 using a stirred undivided BDD/air-diffusion tank reactor at $j$ = 20 mA cm$^{-2}$.

Fig. 6b depicts the pseudo-first-order kinetic analysis of the concentrations determined in the above assays. From the good linear representations found, the $k_1$ values obtained are given in Table 1.

The kinetic analyses confirmed that the higher $k_1$ = 1.261 min$^{-1}$ corresponded to the optimum 0.75 mM Fe$^{2+}$, decreasing to $k_1$ = 1.021 min$^{-1}$ for 1.00 mM Fe$^{2+}$ by the action of reaction (14). Similarly, operating with 0.75 mM Fe$^{2+}$, the lower EE/O value of 0.0522 kWh m$^{-3}$ h$^{-1}$ was achieved.

A comparable outcome is evident in Fig. 7a for the TOC abatement. The percent of this parameter at 180 min of treatment grew progressively from 43% without a catalyst to 77% for 0.75 mM Fe$^{2+}$, whereas it decreased to 75% for 1.00 mM Fe$^{2+}$. Similarly, the higher MCE value of 4.2% was found for the most effective mineralization with 0.75 mM Fe$^{2+}$, as shown in Fig. 7b and Table 1. An anomalous tendency was found for the ECO-H$_2$O$_2$ in which the MCE value slightly increased with prolonged electrolysis because of the efficient mineralization of the most easily oxidizable by-products formed. In contrast, the EC$_{TOC}$ showed a different behavior as a result of the influence of the progressive decay of $E_{cell}$ with the greater amount of catalyst in solution. Fig. 7c shows the increase of EC$_{TOC}$ with time and its gradual decay with raising Fe$^{2+}$ content, achieving a minimum value at 180 min of 1.240 kWh (g TOC)$^{-1}$ for 1.00 mM Fe$^{2+}$, lower than 1.374 kWh (g TOC)$^{-1}$ obtained for 0.75 mM Fe$^{2+}$. The faster degradation/mineralization achieved with 0.75 mM Fe$^{2+}$ was then accompanied by a more expensive process than for 1.00 mM Fe$^{2+}$ by its higher $E_{cell}$ value.

*3.5. Understanding the effect of BP-4 concentration on treatment performance*

EF treatment efficacy varies with contaminant concentration levels and its ability to react with generated ROS. The impact of this variable was examined under the optimized conditions of 0.75 mM Fe$^{2+}$, pH = 3.0, and $j$ = 20 mA cm$^{-2}$. Fig. 8a highlights that the rate of BP-4 degradation decreased progressively with decreasing its concentration from 40 to 1.0 mg L$^{-1}$, with a shorter time for overall removal from 7 to 2 min. This was reflected in the $k_1$ values (Fig. 8b) since the lower $k_1$ = 0.641 min$^{-1}$ for 40 mg L$^{-1}$ progressively raised to $k_1$ = 2.268 min$^{-1}$ for 1 mg L$^{-1}$ (see Table 1). This trend yielded a decrease in EE/O at lower pollutant concentrations from 0.1021 to 0.0288 kWh m$^{-3}$ order$^{-1}$ when passing from 40 to 1 mg L$^{-1}$ (see Table 1), with an insignificant change in $E_{cell}$.

The above findings point to a best oxidation power of the EF treatment at lower BP-4 content. The observed trend occurs due to limited organic compounds available to react with sustained ROS amounts at the same $j$ value. However, more relevant information can be achieved by analyzing the amount of pollutant removed at a given electrolysis time. Thus, at 2 min, when 1 mg L$^{-1}$ BP-4 was completely removed, its concentration more largely dropped down in 2.48, 4.81, 9.24, 16,75, and 26.25 mg L$^{-1}$ for initial 2.5, 5.0, 10, 20, and 40 mg L$^{-1}$. This is indicative of a greater quantity of oxidized pollutant as its concentration increases, and consequently, it possesses more oxidation ability of the organic matter. This pattern emerges due to the dominant influence of the reaction of more BP-4 molecules with greater amounts of reactive ROS that proceeded from the deceleration of the undesired ROS parasitic reactions. This can be explained by the preferential and faster reaction of ROS with organic compounds in solution (e.g., BP-4) than with inorganic species and/or other ROS.

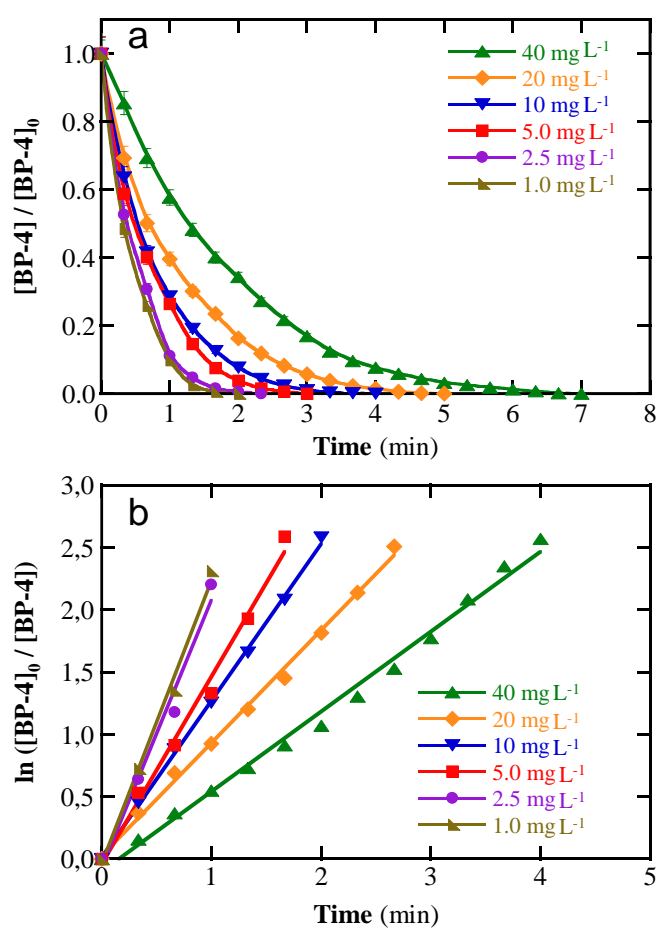

**Fig. 8.** Influence of BP-4 concentration over (a) the change of normalized pollutant content with time and (b) its pseudo-first-order kinetic analysis for the EF treatment of 150 mL of pollutant solutions in ultra-pure water with 0.050 M Na$_2$SO$_4$ and 0.75 mM Fe$^{2+}$ at pH = 3.0 using a stirred undivided BDD/air-diffusion tank reactor at $j$ = 20 mA cm$^{-2}$.

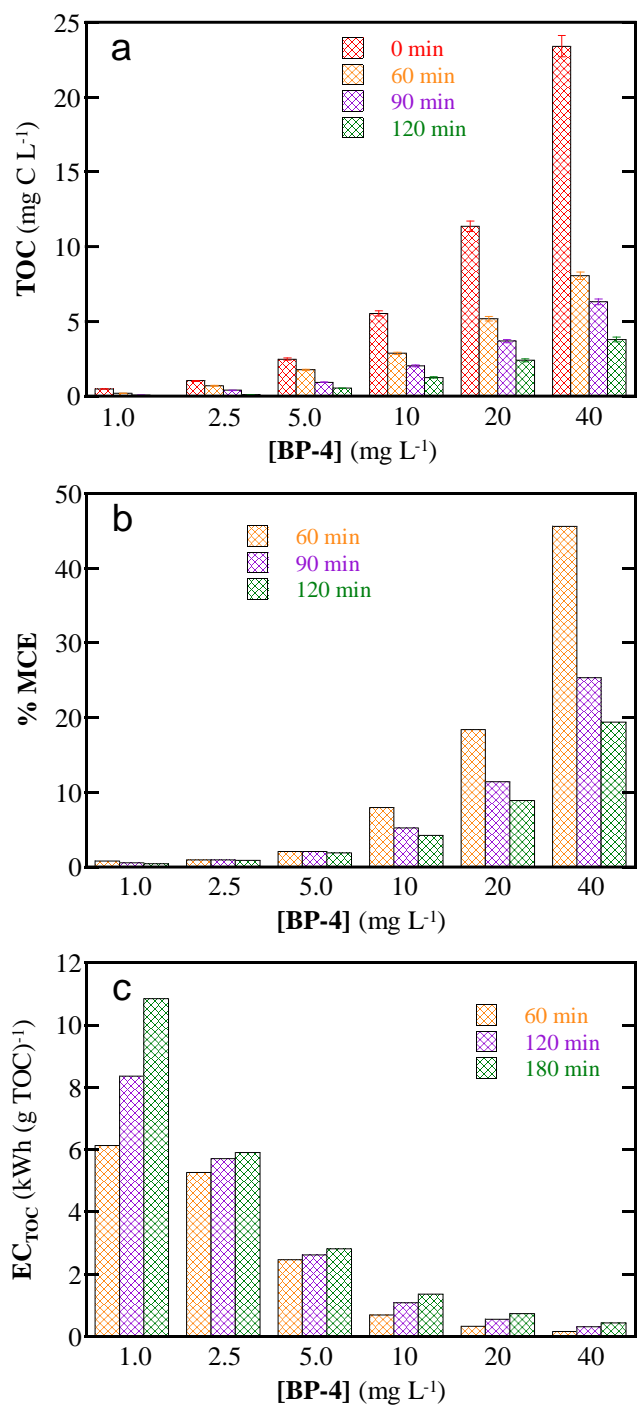

**Fig. 9.** Effect of BP-4 concentration over the variation of (a) TOC, (b) MCE, and (c) $EC_{TOC}$ for the homogeneous EF treatment of 150 mL of pollutant solutions in ultra-pure water with 0.050 M $Na_2SO_4$ and 0.75 mM $Fe^{2+}$ at pH = 3.0 using a stirred undivided BDD/air-diffusion tank reactor at $j$ = 20 mA $cm^{-2}$.

From the above findings, a coherent higher amount of TOC reduction with raising BP-4 concentration is seen in Fig. 9a. While complete mineralization was attained in 180 min at 1.0 mg L$^{-1}$ with a loss of 0.48 mg L$^{-1}$ of TOC, 20.72 mg C L$^{-1}$ of TOC (88% reduction) was removed in 180 min starting from 40 mg L$^{-1}$ pollutant with 23.41 mg C L$^{-1}$ TOC. A minimum TOC decay of 77% was obtained for 10 mg L$^{-1}$ pollutant with 5.54 mg C L$^{-1}$ TOC at 180 min (see Table 1). These findings demonstrate significant improvement in the mineralization system for greater organic load due to the faster reaction of more amount of ROS proceeding from the inhibition of their parasitic reactions. This phenomenon is reflected in the increase in MCE with increasing BP-4 content shown in Fig. 9b and the decay of EC$_{TOC}$ depicted in Fig. 9c. The higher MCE = 45.6% was found at 60 min with 40 mg L$^{-1}$, whereupon it decreased up to 19.4% at 180 min. The lower EC$_{TOC}$ values were determined as well, varying from 0.171 to 0.441 kWh (g TOC)$^{-1}$.

*3.6. Identification of oxidizing agents involved on BP-4 degradation*

The oxidants generated by the homogeneous EF system were identified by specific scavengers like *tert*-butyl alcohol (TBA) quenching •OH and *p*-benzoquinone (*p*-BQ) quenching hydroperoxide radical/superoxide HO$_2$•/O$_2$•$^-$ (He et al., 2019; Mou et al., 2025). Fig. 10 highlights that 10 mM TBA significantly inhibits the oxidation of 10 mg L$^{-1}$ BP-4 under optimum conditions, reducing its removal from 100% to 8% after 4 min of electrolysis. A lower reduction of up to 42% can be seen for 1 mM TBA. In contrast, for 10 mM *p*-BQ, the content of BP-4 slightly decayed from 100% to 99% at the same time, with scarce effect of 1 mM *p*-BQ. These findings confirm that BDD(•OH) and •OH are the main ROS for BP-4 degradation/mineralization in the EF system, without notable contribution of HO$_2$•/O$_2$•$^-$.

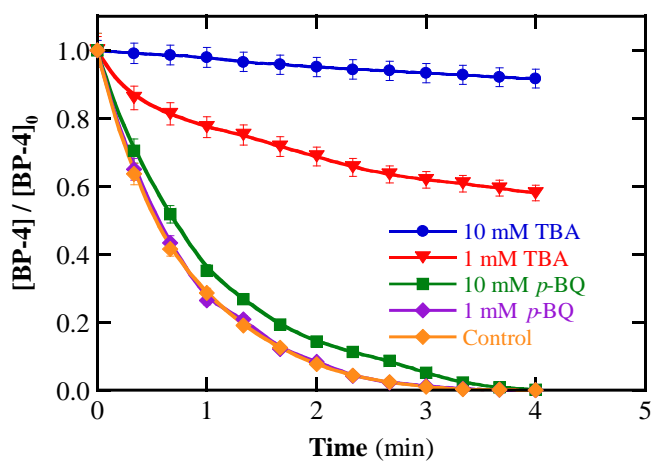

**Fig 10**. Effect of 1 and 10 mM *tert*-butanol (TBA) and 1 and 10 mM *p*-benzoquinone (*p*-BQ) as scavengers over the change of normalized BP-4 content with electrolysis time for the homogeneous EF process of 150 mL of 10 mg L$^{-1}$ pollutant with 0.050 M Na$_2$SO$_4$ and 0.75 mM Fe$^{2+}$ at pH = 3.0 using a stirred undivided BDD/air-diffusion tank reactor at $j$ = 20 mA cm$^{-2}$.

*3.7. Assessing the evolution of carboxylic acids as main by-products*

To better understand the mineralization process of BP-4 by EF, final carboxylic acids including oxalic, formic, acetic, and fumaric were detected by ion-exclusion HPLC and their evolution was determined during the electrolysis of 10 mg L$^{-1}$ pollutant with 0.75 mM Fe$^{2+}$ at pH = 3.0 and $j$ = 20 mA cm$^{-2}$. The two latter acids are produced from the breaking of the benzene moieties of the aromatic intermediates. They are subsequently oxidized to the ultimate formic and oxalic acids that directly give CO$_2$ (Brillas et al., 2009; Ganiyu et al., 2021; Brillas, 2023). Fig. 11a shows a growth of the concentration of such acids in the order: fumaric acid < acetic acid < formic acid << oxalic acid. While the content of the first three acids remained more or less steady from 30 min close to 0.07, 0.17 and 0.22 mg L$^{-1}$, respectively, oxalic acid attained about 1.9 mg L$^{-1}$ between 45 and 90 min, decreasing to 1.0 mg L$^{-1}$ after 180 min. These by-products persisted during electrolysis because they formed very stable Fe(III) complexes that are slowly destroyed by oxidants BDD(•OH) and •OH (Brillas et al., 2009).

Fig. 11b shows the evolution of the TOC components of BP-4/aromatic by-products and the 4 carboxylic acids with electrolysis time. It can be seen the predominance of carboxylic acids formed from 45 min, so that, only these by-products are present in solution at 180 min. It is important to remark that these low molecular weight carboxylic acids are highly biodegradable. This persistence of the final Fe(III)-carboxylate species explains that no overall mineralization can be achieved by the EF process, attaining a maximum of 88% mineralization in the different processes considered.

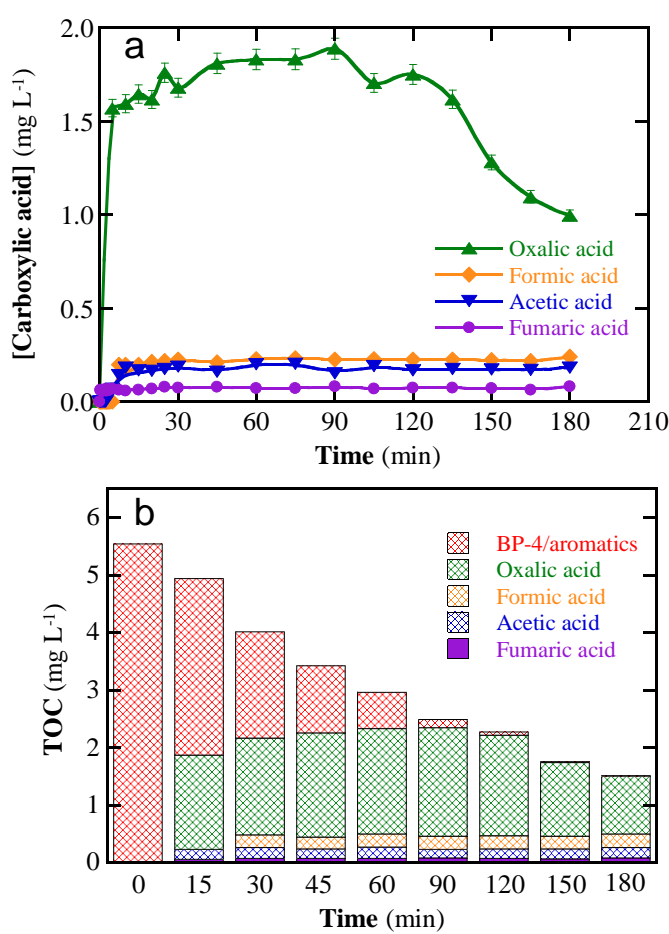

**Fig. 11.** Time course of (a) the concentration of final carboxylic acids formed and (b) the related TOC including that of BP-4/aromatic by-products determined for the homogeneous EF treatment of 150 mL of pollutant solutions in ultra-pure water with 0.050 M $Na_2SO_4$ and 0.75 mM $Fe^{2+}$ at pH = 3.0 using a stirred undivided BDD/air-diffusion tank reactor at $j$ = 20 mA cm$^{-2}$.

*3.8. Operational longevity and reproducibility of the homogeneous EF process*

The ability of the homogeneous EF system to maintain consistent performance across multiple treatment cycles is critical for practical implementation. To evaluate this, 10 consecutive runs of 4 min each were performed with 150 mL of 10 mg L$^{-1}$ BP-4 in 0.050 M Na$_2$SO$_4$ using optimized conditions (0.75 mM Fe$^{2+}$, pH 3.0, and $j$ = 20 mA cm$^{-2}$). Fig. 12 shows the complete degradation achieved in all 10 cycles following a similar decay within the 4 min treatment demonstrating an excellent reproducibility of the process. No Fe$^{3+}$ precipitation was observed. These results confirm the stable operation of the system under repeated use, with no observable decline in efficiency. This consistency suggests that key components, including the oxidative capacity of BDD sustained H$_2$O$_2$ electrogeneration at the carbon cloth air-diffusion cathode, and efficient Fe$^{2+}$/Fe$^{3+}$ cycling, remain fully functional without electrode fouling or deactivation (Yu et al., 2022).

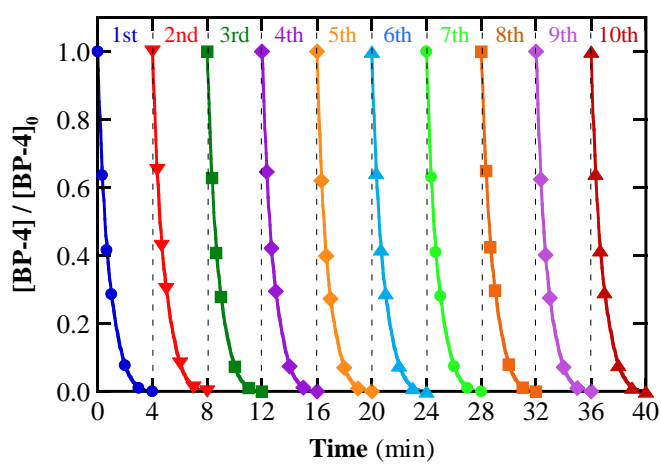

**Fig. 12.** Reusability of the system in ten consecutive degradation runs of homogeneous EF of 150 mL of pollutant solutions in ultra-pure water with 0.050 M Na$_2$SO$_4$ and 0.75 mM Fe$^{2+}$ at pH = 3.0 using a stirred undivided BDD/air-diffusion tank reactor at $j$ = 20 mA cm$^{-2}$.

## 4. Conclusions

This study shows that homogeneous EF is able to rapidly and efficiently degrade/mineralize BP-4 solutions in ultrapure water with a BDD/air-diffusion cell. The optimum pH for the removal of the pollutant was 3.0, as expected if the main oxidant was homogeneous $^{\bullet}$OH formed from Fenton reaction. Homogeneous EF was considerably more effective than AO-H$_2$O$_2$ treatment due to the high oxidation capability of generated $^{\bullet}$OH. The procedure has been optimized at this pH determining the effect of $j$, Fe$^{2+}$ dosage, and pollutant content. The BP-4 concentration abatement obeyed a pseudo-first-order reaction. The best operating conditions have been found for $j = 20$ mA cm$^{-2}$, 0.75 mM Fe$^{2+}$, and 40 mg L$^{-1}$ pollutant where it totally disappeared in 7 min with $k_1 = 0.641$ min$^{-1}$ and EE/O = 0.1021 kWh m$^{-3}$ order$^{-1}$ and its TOC was reduced by 88% with MCE = 19.4% and EC$_{TOC}$ = 0.261 kWh (g TOC)$^{-1}$ after 180 min. The process became very effective even for a low BP-4 concentration of 1.0 mg L$^{-1}$, achieving overall contaminant decay in 2 min and complete mineralization in 180 min, but with much lower MCE and EC$_{TOC}$ due to the consumption of generated ROS by parasitic reactions. The large hindering found upon addition of TBA confirmed that BDD($^{\bullet}$OH) and $^{\bullet}$OH were the preferent oxidants of organics. Fumaric, acetic, formic, and oxalic acids have been found as final carboxylic acids. After 180 min of electrolysis, these acids were the only by-products that persisted in the treated solution because they originated recalcitrant Fe$^{3+}$ complexes that are hardly oxidized by BDD($^{\bullet}$OH) and $^{\bullet}$OH. In summary, the homogeneous EF process offers several advantages for BP-4 degradation, including *in-situ* H$_2$O$_2$ generation, high mineralization efficiency, and operation at mild conditions. The system offers direct applicability for industrial wastewater and municipal treatments plants, with advantages like sludge-free operation, fast treatment, and cycle reproducibility and stability. Compared to conventional Fenton methods, this electrochemical method produces continuous H$_2$O$_2$ while maintaining excellent degradation kinetics. However, challenges such as pH dependence (optimal at ~3) and energy consumption must be considered. When benchmarked against alternatives like photocatalysis or ozonation, homogeneous EF often demonstrates superior mineralization rates and lower operational costs for acidic wastewater, though its scalability may be limited by electrode materials and reactor design.

**Acknowledgements**

The authors are grateful to Fundação de Amparo à Pesquisa do Estado de São Paulo (FAPESP, #2022/10484-4, #2024/03549-8, #2022/15252-4, #2022/12895-1) and Conselho Nacional de Desenvolvimento Científico e Tecnológico (CNPq, #308663/2023–3, #402609/2023–9) for the financial support.